# Ultrashort laser-induced nuclear reactions: initiating decay of helium nuclei and subsequent fusion reactions.

**L. M. Kovachev[1*], E. Iordanova[2], G. Yankov[2], I. P. Angelov[1]**

[1]*Institute of Electronics, Bulgarian Academy of Sciences, 72 Tzarigradsco chausse, 1784 Sofia, Bulgaria*
[2]*Institute of Solid State Physics, Bulgarian Academy of Sciences, 72 Tzarigradsco shoussee, 1784 Sofia, Bulgaria*
*[*lubomirkovach@yahoo.com](mailto:lubomirkovach@yahoo.com)*

**Abstract:** We present a novel method to construct particle accelerators targeting light atoms and nuclei using high-power femtosecond laser pulses. Initially, we confine light atoms within the laser pulse envelope due to longitudinal polarization forces, allowing them to acquire kinetic energies of several GeV. Subsequently, an external electric field separates the nuclei at the cathode, concentrating helium nuclei in a small area. The kinetic energy of the 1.88 GeV impacts, exceeding the alpha particle binding energy (28 MeV) by two orders of magnitude, induces powerful gamma radiation and neutron emission from decay processes. This experiment marks a demonstration of a laser-induced decay method for helium nuclei for the first time. Moreover, helium isotopes or deuterium nuclei trapped on the cathode show significantly reduced Coulomb repulsion, enabling subsequent nuclear fusion reactions and substantial nuclear energy release.

## 1. Introduction

The polarization force from lasers operating in continuous wave (CW) mode was first identified in earlier studies [1-2], where it was shown to be proportional to the transverse gradient of the square of the electric field. Three years later, Gordon [3] introduced an additional force term for pulsed laser regime associated with the Poynting vector and the energy flow. Gordon suggested that this longitudinal polarization force repels neutral atoms from the laser pulse. However, in our subsequent studies [4, 5, 6], we demonstrated that this force is attractive, drawing particles towards the center of the laser pulse, with the pulse intensity acting as an attractive potential. This attractive longitudinal force becomes significant in the femtosecond regime, inversely proportional to the pulse duration.

The main finding of these studies is the demonstrated conditions under which neutral particles can be confined within the pulse envelope, allowing femtosecond laser pulses to accelerate neutral particles to energies of several GeV.

A significant challenge in using this laser method for nuclear collisions or fusion is removing electron envelopes from atoms and compressing nuclei into a confined space. Previous work [7] showed this could be achieved using an external electric field in a cylindrical condenser. In this paper, we present a novel method where collisions on the cathode, with energies of several GeV and the accumulation of alpha particles in a localized area, lead to a new laser-induced decay method for helium nuclei. Additionally, we discuss the potential for subsequent nuclear fusion reactions between helium isotopes or deuterium nuclei, resulting in the release of significant nuclear energy with applications for fusion reactors.

## 2. Femtosecond laser pulse as an accelerator of neutral particles.

*2.1 Theoretical background*

In this section, we briefly present the theory of the polarization force in the context of a linearly polarized laser pulse with an amplitude envelope $\vec{A} = A_x \vec{x}$ extending up to a few meters from the source. The nonlinear polarization density force $\vec{f}^P$ of a laser pulse in isotropic media can be expressed as [5, 7]:

$$\vec{f}^P(x,y,t) = \frac{4\pi}{c^2} \frac{d\left[\left(\chi^{(1)} + \chi^{(3)}|\vec{E}|^2\right)\vec{S}\right]}{dt}, \qquad (1)$$

where $c$ is the speed of light in vacuum, $\chi^{(1)}$ and $\chi^{(3)}$ are the linear and nonlinear susceptibilities of the media, $|\vec{E}|^2 = |\vec{A}|^2$ is the square modulus of the electric field, and $\vec{A}$ is the amplitude envelope of the laser pulse. The nonlinear addition to the force is three orders of magnitude less than the linear part $\chi^{(3)}|\vec{E}|^2 \ll \chi^{(1)}$, as can be estimated from Equation (1); thus, we focus on the influence of the linear part only. The optical propagation scale of a laser pulse in the linear regime is characterized by diffraction and dispersion lengths. For a typical 30-35 fs laser pulse, diffraction and dispersion lengths vary widely, ranging from 40-50 m. The spot and its temporal profile do not change significantly at distances of 5-6 meters, where our experiments were conducted. Therefore, we represent the amplitude envelope of a linearly polarized Gaussian pulse in the first-order dispersion approximation in Galilean coordinates $z = z' - v_{gr} t'; t = t'$ as:

$$A_x(x,y,z) = A_0 \exp\left(-\frac{x^2 + y^2}{2d_0^2} - \frac{z^2}{2z_0^2}\right), \qquad (2)$$

where $d_0$ is the initial spot of the laser pulse and $z_0$ is its longitudinal spatial width.

It is important to point here that the dynamical properties of the optical force (1) and the applied potential at long distances, due to diffraction, and dispersion, request firstly to obtain exact solutions of the corresponding amplitude equation [8, 9].

Rewriting the linear part of the polarization force Eq (1) in the exact coordinates and using the fact that the Poynting vector approximation on these distances is $\vec{S} = S_z \vec{z} = I\vec{z} = \left[n_0 c |A_x|^2 / 2\pi\right]\vec{z}$ as well as the facts that $I_0 = n_0 c |A_x|^2 / 2\pi$ and $dt = dz/v_{gr}$ we obtain

$$\vec{f}^P(x,y,z) = f_z(x,y,z)\vec{z} = \frac{2n_0 \chi^{(1)} v_{gr}}{c} \frac{d|A_x|^2}{dz}\vec{z}. \qquad (3)$$

Substituting the solution of the laser field Eq (2) into the expression for the density force, Eq (3), we find:

$$\vec{f}^{\,P} = f_z^P(x,y,z)\vec{z} = -I_0 \frac{8\pi\chi^{(1)} v_{gr}}{c^2} \frac{z}{z_0^2} \exp\left(-\frac{x^2+y^2}{d_0^2}\right)\exp\left(-\frac{z^2}{z_0^2}\right)\vec{z}. \quad (4)$$

The force is negative, with the pulse front attracting particles toward the center while the back side also pushes them toward the center. The force density is a gradient in the $z$-direction, and a potential density can be introduced by:

$$dU(x,y,z) = \vec{f}^{\,P} \cdot d\vec{z} = -f_z^P(x,y,z)dz. \quad (5)$$

The sign minus is used because we have *negative work* when the force is in the opposite direction on the translation vector $d\vec{z} = |dz|\vec{z}$. Given that the force depends on 3-dimensional coordinates, the potential density can be naturally introduced by integrating Equation (5) and using the relation $I_0 = n_0 c |A_x|^2 / 2\pi$:

$$U(x,y,z) = -I_0 \frac{4\pi\chi^{(1)} v_{gr}}{c^2} \exp\left(-\frac{x^2+y^2}{d_0^2}\right)\exp\left(-\frac{z^2}{z_0^2}\right) \quad (6)$$

The pulse's Gaussian spatial shape acts as an attractive potential, and Fig. 1 shows a graph of the potential density.

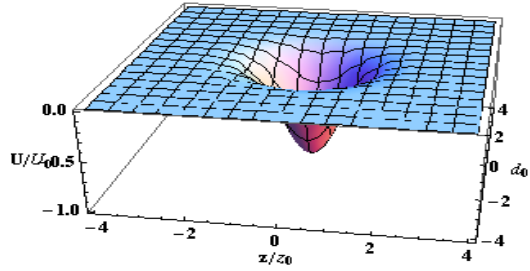

Fig. 1. Graphic of Gaussian laser pulse potential density. The pulse's shape moves with the group velocity and plays the role of an attractive potential.

The formulae for the force density Eq (3) and the potential density Eq (6) are integrated over the whole space to obtain accurate, measurable forces from the density ones. After integration, we obtained two constants related to the square of the pulse diameter $d_0^2$ and the longitudinal size $z_0 = v_{gr} t_0$. The effective real force at the peak of the pulse at level $e^{-1}$, expressed in conventional units, is:

$$\langle F \rangle_z^P = -8\pi^{5/2} \chi^{(1)} \frac{v_{gr}^2}{c^2} \frac{E_0}{z_0} = -8\pi^{5/2} \chi^{(1)} \frac{v_{gr}}{c^2} \frac{E_0}{t_0}, \quad (7)$$

and for the potential is obtained:

$$\langle U \rangle_z^P = -4\pi^{5/2} \frac{v_{gr}^2}{c^2} \chi^{(1)} E_0, \quad (8)$$

where $E_0$ is the initial energy of the laser pulse. As a gradient force, the optical force is inversely proportional to the laser pulse's longitudinal length or temporal duration. Long pulses exhibit relatively small gradients compared to femtosecond ones, making this force negligible for longer pulses. An important question arises: What is the depth of the radiation potential in gases? To address this, we compare it to the Boltzmann energy of free particles at room temperature

$$U_B = k_B T = 4.14 \times 10^{-21} \quad [J]. \tag{9}$$

The depth of the radiation potential in gases can be compared to the Boltzmann energy of free particles at room temperature $T = 300\ K$. For a 35-fs laser pulse and initial energy $E_0^{laser} \cong 7\ mJ$, as observed in our experimental investigation, the potential is in the range of $10^{-5} - 10^{-7}$:

$$U_z^{eff} \cong 1.\times 10^{-5} - 1.\times 10^{-7} \quad [J]. \tag{10}$$

This potential is fourteen orders of magnitude greater than the Boltzmann energy and indicates that the attractiveness of the femtosecond laser pulse, as described by Eq. (8), is sufficiently strong to confine and cool particles within the pulse envelope, allowing them to gain significant kinetic energy and move at the group velocity of the pulse. What new physical processes can be observed following this confinement of particles?

The confinement of particles within the pulse envelope can lead to several new physical processes:

1. The probability of collision with free atoms and molecules in the air increases significantly due to the high density of trapped particles entering the pulse's spot. The collision energies, ranging from 12–24 GeV, exceed the ionization energy of neutral atoms, suggesting a new regime of collision ionization with femtosecond pulses at low intensities. The main distinction from the well-known multi-photon and tunnel ionization processes is that this new regime can be observed at low intensities, on the order of $I \approx 10^{10-12}\ W/cm^2$. Ionization with femtosecond pulses at such low intensities during filamentation was recently observed in [10].

2. In the nonlinear regime, neutral moving particles do not generate third harmonics but instead emit at a frequency proportional to three times the carrier-envelope frequency $3\omega_{THz} = 3k_0(v_{ph} - v_{gr})$. This coherent generation has been experimentally observed in air [11, 12].

3. Neutral atoms (e.g., helium) confined within the pulse envelope acquire kinetic energy around $E_{He}$~1.88 GeV, which is two orders of magnitude greater than the binding energy of alpha particles $E_\alpha$~28 MeV. This makes them capable of undergoing nuclear reactions, such as the decay of helium nuclei to helium isotopes and deuterium.

To facilitate this nuclear reaction, a new approach must be applied to remove the electron shell of helium atoms, converting them into alpha particles while retaining their kinetic energy of motion of 1.88 GeV. This novel methodology enables the realization of a unique laser fusion scheme, distinct from the well-known methods such as laser-plasma driven and magnetic confined plasmas (e.g., Tokamak). In this way, the femtosecond lasers with a combination of such method can be transformed into the cheapest devices for particle acceleration. We developed such a prototyped device [7], and in the following section, we discuss experiments observing nuclear reactions that lead to powerful gamma radiation and neutron scattering.

### 2.2 Experimental setup

The schematic representation of the experimental setup for capturing and accelerating helium particles with laser pulses, followed by the separation of their nuclei on the cathode to create an electrostatic field, conversion into nuclear isotopes, and subsequent thermonuclear fusion, is presented in Fig. 2. The setup consisted of a femtosecond laser and a quartz vacuum tube aligned along the optical axis. A transverse helium flow system was attached to the first flange of the vacuum tube. One side of the first flange was connected to a gas bottle filled with helium, which was connected to the vacuum tube via a pressure-reducing valve to control the helium gas flow. The opposite side of this flange was connected to vacuum pump 1, which pumped the directed flow perpendicular to the optical axis. Quartz windows were attached to the first and second flanges. Vacuum pump 2 was connected to the second flange to manipulate the vacuum within the tube. A stainless-steel anode tube was placed in the middle of the vacuum tube, with a tungsten round wire (6 mm diameter) cathode mounted in its center, connected to the negative exit of a high-voltage (HV) generator providing a constant voltage of up to 25 kV. High-sensitivity GR-110G radiation detector (FUGRO INSTRUMENTS, Australia) and neutron detectors were positioned opposite the middle section of the vacuum tube.

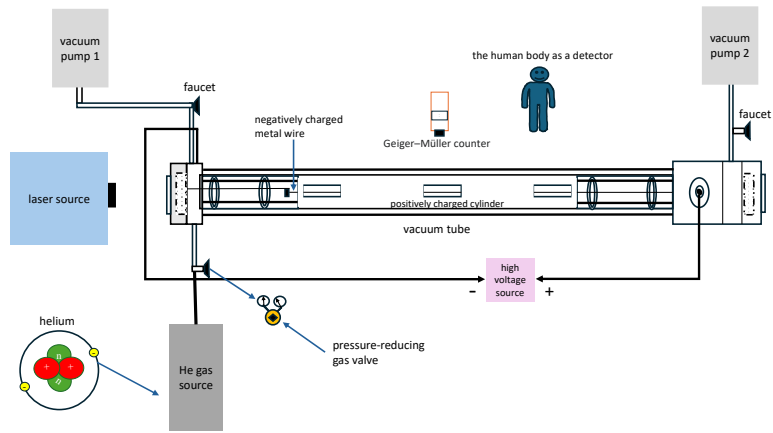

Fig. 2. Schematic view of the setup configuration for capturing and accelerating helium particles with laser pulses, with subsequent separation of their nuclei on the cathode wire of a system for creating an electrostatic electric field and their conversion into nuclear isotopes.

In the experiments, the laser source was a 35-fs laser operating at a repetition rate of 1 kHz, with an initial pulse energy $E_0 \sim 6\ mJ$ and a spot radius at $e^{-1}$ level $d_0 = 3\ mm$. Our main request is to work in a regime slightly below plasma generation, where the polarization of the particles obtained from the laser radiation is maximal. This requirement follows from the fact that the force depends on the polarization of the particles [5, 7].

### 3. Experimental results

Initially, a vacuum was created in the vacuum tube. Helium atoms were introduced through the flange at the beginning of the vacuum tube, orthogonal to the laser's optical path, achieving a transverse channel with a pressure range of 0.2 to 0.5 atmosphere using a combination of a gas bottle, pressure-reducing valve, and vacuum pump. A DC voltage of 10 kV was applied to the anode tube and cathode wire. When femtosecond laser pulses (dimeter on level $e^{-1}$ is ~6 mm) were directed parallel to the cathode at distance 0.5-0.7 mm from it, an intense light spot with size of 1-2 cm was observed on the tungsten cathode wire, 40 cm from the beginning of the electrode configuration. Fig 3 presents the schematic picture of this result. Given the presence of only transverse helium flow in the vacuum tube, the light spot was attributed to positively charged alpha particles hitting the cathode. This conclusion was supported by the absence of coronal or Townsend discharge around the cathode, indicating the light spot was localized on the cathode. The polarization force confined helium particles within the laser pulse, dissipating once the particles were ionized and converted into alpha particles in the DC electric field. These particles retained the high horizontal group velocity of the laser pulse, left the pulse envelope, and impacted the cathode with approximately 1.88 GeV of energy, following a parabolic trajectory (Fig. 3).

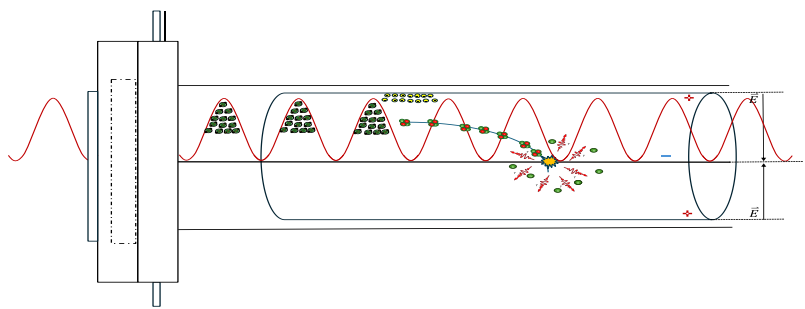

Fig. 3. Schematic view of the observed in the experiment light spot on the cathode wire at 40 cm from the beginning of the electrode configuration on the tungsten cathode. The assumed trajectory of the alpha particles upon their impact on the cathode with an energy of 1.88 GeV is plotted.

It is well established that the binding energy of the nuclides in an alpha particle is $U_{nucl}=28$ MeV. However, the kinetic energy of their collisions with the cathode is significantly higher, approximately $U_{kin}=1.88$ GeV, which is much greater than $U_{nucl}$. Consequently, an experimental demonstration of helium particle capture and the subsequent acquisition of substantial kinetic energy by alpha particles, on the order of several GeV, would be evidenced by the observation of nuclear decay of alpha particles accompanied by the emission of gamma radiation and free neutrons into the surrounding environment.

Anticipating this phenomenon, we positioned the highly-sensitive GR-110G radiation detector directly opposite to the observed light spot (refer to Fig. 2). The experimental results exceeded our expectations. Adjusting the working conditions such as, gas flow, laser pulses, voltage level and following the activation of the laser source, the detector rapidly reached its maximum level, recording values 28 times greater than the baseline levels. Figure 4 presents from the experiment the detected gamma radiation in particles per second. The minima in radiation observed in the graph correspond to intervals when the laser pulses were obstructed by a metal plate. This correlation confirms that the observed effect is attributable to laser confinement, enabling the particles to achieve the high energies necessary for the nuclear decay of alpha particles.

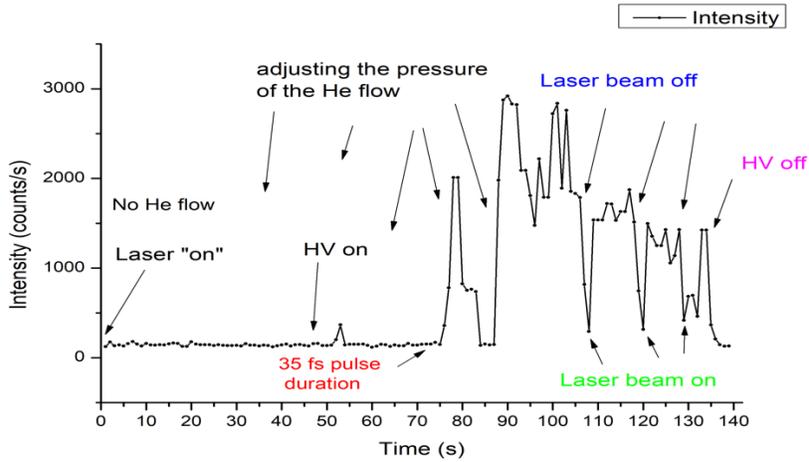

Fig 4. Gamma radiation was detected in particles per second from the GR-110G radiation detector experiment. The minimums in radiation on the graph correspond to the moments when the laser pulses are stopped by a metal plate. The maximums correspond to radiation 28 times greater than the ground state.

As we do not expected so high level of radiation, medical examinations of the experimental staff were conducted to detect gamma radiation exposure and observe damage from heavy particles such as free neutrons. Immediately after the experiment, we visited the National Center for Radiobiology and Radiation Protection [13]. It was determined that during one minute of work in the radiation zone (see Fig. 4), we received an average dose of 300 mSi of gamma radiation. Furthermore, significant DNA damage from free neutrons was detected, with up to 5 damaged DNA strands out of 500 tested. Figure 5 presents the results of these investigations [14], showing the measured radiation dose from gamma radiation and the DNA damage caused by heavy particles such as neutrons.

## Dicentrics dose estimation report

Biodose Tools v3.6.1

22 May, 2024 07:56:11

**Curve used**

**Fit formula**

$$Y = C + \alpha D + \beta D^2$$

**Coefficients**

| | estimate | std.error | statistic | p.value |
|---|---|---|---|---|
| $C$ | 4.080e-04 | 6.618e-04 | 6.165e-01 | 1.129e+00 |
| $\alpha$ | 1.238e-02 | 8.565e-03 | 1.445e+00 | 4.161e-01 |
| $\beta$ | 7.493e-02 | 5.257e-03 | 1.425e+01 | 6.128e-05 |

**Case data analyzed**

| $N$ | $X$ | $C_0$ | $C_1$ | $C_2$ | $C_3$ | $C_4$ | $C_5$ | $y$ | $\hat{\sigma}_y/\sqrt{N}$ | $\hat{\sigma}^2/\bar{y}$ | $u$ |
|---|---|---|---|---|---|---|---|---|---|---|---|
| 500 | 5 | 495 | 5 | 0 | 0 | 0 | 0 | 0.010 | 0.004 | 0.992 | -0.142 |

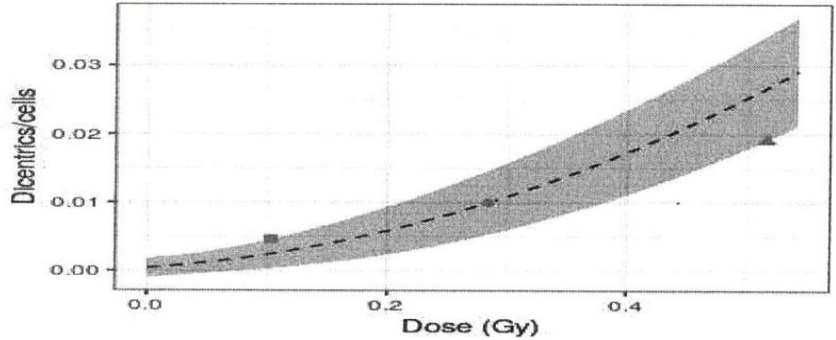

Fig. 5. The table represents the damages to our DNA by free neutrons: up to 5 damaged DNA from 500 tested. These damages are obtained for a one-minute exposition of the body of one person of the experimental staff. The radiation dose from this exposition is plotted on the graph.

These findings underscore the critical importance of robust radiological protection during such experiments. A principal conclusion from this study is the imperative: Do not conduct such experiments without substantial radiological safeguards.

What types of nuclear decay and fusion reactions could result in such significant emissions? The massive emission observed suggested two possible decay channels for the helium nucleus. The first involves the decay of a helium nucleus into a helium isotope, accompanied by the release of a neutron and high-energy gamma particles. The second channel entails the decay of a helium nucleus into two deuterium nuclei—hydrogen isotopes—also with the emission of high-energy gamma particles:

$$\text{I. } He^4 \rightarrow He^3 + n + \gamma$$

$$\text{II. } He^4 \rightarrow H_1^2 + H_1^2 + \gamma.$$

An important feature of cathode-trapped helium isotopes is their accumulation in one place, significantly reducing the Coulomb barrier between nuclei due to the electric field of the cathode. Overcoming this barrier enables secondary fusion reactions of helium and hydrogen isotopes along the following channels:

$$\text{I. } He^3 + He^3 \rightarrow He^4 + p + p + \gamma = 12.86 \text{ MeV}$$

$$\text{II. } He^3 + H_1^2 \rightarrow He^4 + n + \gamma = 18.3 \text{ MeV}$$

$$\text{III. } H_1^2 + H_1^2 \rightarrow He^3 + n + \gamma = 4 \text{ MeV}.$$

## 4. Discussion

Previous studies [4-6] have discussed the confinement of neutral particles within a pulse envelope, emphasizing their applications in GHz generation and collision ionization during femtosecond laser pulse filamentation in air. We have demonstrated that accelerated neutral atoms can be compressed by nonlinear optical mechanisms to distances up to several times the electron orbitals size around the nuclei. This insight led us to develop a method and construct a device designed *to separates nuclei from electrons while preserving their kinetic energy at several* GeV. The constructed device allows the nuclei with the substantial energy to be compressed and collected onto the cathode zone of the device.

Given that the collision energy exceeds the binding energy of alpha particles by two orders of magnitude, our experiment successfully observed the nuclear decay of alpha particles, accompanied by the release of gamma radiation and free neutrons into the surrounding space. The presence of radiation and free neutrons was confirmed using a gamma detector and further validated by radiological analysis, which measured the radiation dose in the bodies of the research team. Furthermore, the detected significant DNA damage from free neutrons, is evidence of a decay of helium nuclei to nuclei of helium isotopes and deuterium.

The achieved conditions demonstrate that the helium and nitrogen isotopes confined onto the cathode zone. Considering the significantly reduced Coulomb repulsion between the confined nuclei, suggests the high probability of secondary fusion reactions. These findings reveal a novel mechanism with significant implications for fundamental nuclear physics and practical applications, underscoring the innovation and importance of our research.

The novelty of our research lies in demonstrating that fs laser pulses can effectively accelerate neutral particles and light nuclei to unprecedented energies, facilitating nuclear decay and potential secondary fusion reactions. This advancement has remarkable implications for society and the economy, opening new pathways for developing laser-based nuclear reactors providing a cleaner, more efficient energy source, and addressing global energy demands while reducing reliance on fossil fuels. Additionally, these findings pave the way for further research into high-energy particle acceleration and nuclear fusion, potentially leading to breakthroughs in energy production and fundamental physics.

**Funding**

The National Roadmap for Research Infrastructure, Bulgaria "Extreme Light" ELI-ERIC Consortium under contract D01-351/13.12.2023.